\documentclass[aip,amsmath,amssymb,reprint]{revtex4-1}
\usepackage{graphicx}% Include figure files
\usepackage{epsfig} %Works faster than the graphicx package}
\usepackage{dcolumn}% Align table columns on decimal point
\usepackage{bm}% bold math
\usepackage{epstopdf}
\usepackage{multirow}
\usepackage{amsmath}
\usepackage{booktabs}
\usepackage{color}
\usepackage{gensymb}
\usepackage{dblfloatfix}
\usepackage{kantlipsum}  
\usepackage{hyperref}
\usepackage{ulem}

\usepackage[utf8]{inputenc}
\usepackage[T1]{fontenc}
\usepackage{mathptmx}

\begin{document}

\title{Compact 2D Magneto-Optical Trap for Ultracold Atom Setups}

\preprint{APS/123-QED}

\author{A.Z. Lam}
\author{C. Warner}
\author{N. Bigagli}
\author{S. Roschinski}
\author{W. Yuan}
\author{I. Stevenson}
\author{S. Will}
\email[To whom correspondence should be addressed: ]{sebastian.will@columbia.edu}
\affiliation{%
 $^1$Department of Physics, Columbia University, 538 West 120th Street, New York, New York 10027, USA}

\date{\today}% It is always \today, today

\begin{abstract}
We report on the design, implementation, and performance of a compact two-dimensional magneto-optical trap (2D MOT)  for cesium. In a small-volume vacuum chamber, the setup uses cesium dispensers in close proximity to the trapping region of the 2D MOT and operates at low vapor pressures in the $10^{-9}$ torr range. We achieve a cold atom flux of $4 \times 10^8$ atoms/s that is comparable to the performance of more complex atomic sources. The setup is simple to construct and can be adapted to a broad range of atomic species.
%We report on the design, implementation, and performance of a novel compact 2D magneto-optical trap. Our setup features cesium dispensers that are directly integrated into a small-volume vacuum chamber. We observe a peak atom flux of $4 \times 10^8$~atoms/s, which is comparable to the performance of more complex cesium sources. Using a numeric model, we analyze the performance of our setup. Our design can be readily adapted to a broad range of atomic species, making an important contribution to simplification and standardization of ultracold atom setups. 
\end{abstract}
\maketitle

\section{Introduction}

Over the past decades, ultracold atom setups have evolved from proof-of-concept demonstrators to elaborate apparatuses that enable quantum technologies, such as atom interferometry,~\cite{cronin2009optics} precision timekeeping,~\cite{2015YeClocks,Koller2017transportable} and quantum simulation.~\cite{gross2017quantum, browaeys2020many} With the overall complexity of such setups increasing, there is a need for high-flux atomic sources that are easy to use and maintain. 2D MOTs have emerged as a versatile source technology.~\cite{dieckmann1998two, schoser2002intense, chaudhuri2006realization, catani2006intense, ridinger2011large, huang2016intense,kreyer2017cyl} In 2D MOTs, atomic vapor is transversely cooled in two dimensions, generating a cold atomic beam that escapes longitudinally. Compared to Zeeman slowers,~\cite{phillips1982laser, streed2006large, van2007large, reinaudi2012dynamically} which rely on longitudinal laser cooling and require a magnetic field profile designed for specific atomic species, 2D MOTs are technically simpler, can be constructed with a smaller footprint, and easily adapted to a range of atomic species.

%Atomic beams are typically generated from atomic vapors that are created via effusive ovens~\cite{dieckmann1998two, schoser2002intense,tiecke2009high} or dispensers.~\cite{demarco1999enriched,catani2006intense,moore2005collimated, holtkemeier20112d,weber2003bose} Ovens are relatively easy to install in a vacuum chamber, but require a macroscopic amount of atomic raw material, which comes with the risk of ion pump poisoning, coating of viewports, and additional dangers of handling bulk quantities of reactive materials. Dispensers are more compact and only contain a few milligrams of material.  Running an electrical current through a dispenser produces an atom jet on demand, leading to lower partial pressures than an oven and reducing the risk of ion pump poisoning. However, the mounting structure for dispensers  needs to be carefully designed, requiring thermal and electrical isolation from the vacuum system. % and, to operate efficiently, are ideally positioned in line-of-sight to the 2D MOT.One possible solution is to mount the dispensers on a  complex customized flange.~\cite{jollenbeck2011hexapole, dorscher2013creation} In other cases, dispensers are simply used to generate a background vapor much like an oven would,~\cite{weber2003bose, catani2006intense} which comes at the cost of %reduced atomic flux of the 2D MOT or fast depletion of the dispenser reservoir.

Atomic vapors for 2D MOT setups are often generated using effusive ovens that contain a macroscopic amount of atomic raw material and operate at vapor pressures of up to $10^{-7}$ torr in the trapping region.~\cite{dieckmann1998two, schoser2002intense, chaudhuri2006realization, ridinger2011large, huang2016intense}~Such setups can generate high flux atomic beams of up to $10^{10}$ atoms/s,~\cite{schoser2002intense,chaudhuri2006realization,huang2016intense} but the elevated pressures increase the need for regular maintenance. Typical complications include coating of viewports, ion pump poisoning, and hazardous conditions when opening the setup to air. Also, it is challenging to maintain ultrahigh vacuum (UHV) with pressures below $10^{-11}$ torr in attached science chambers, which is often required in modern ultracold atom setups. 

Dispensers are a well-known alternative to generate atomic vapors in a more controlled way.\cite{demarco1999enriched,weber2003bose,moore2005collimated,catani2006intense,holtkemeier20112d,jollenbeck2011hexapole,dorscher2013creation, kreyer2017cyl}~Dispensers typically contain a few milligrams of atomic raw material and produce an atom jet on demand by running an electrical current through them. 2D MOTs with dispensers can be operated in elevated pressure regimes similar to ovens, generating high flux atomic beams~\cite{jollenbeck2011hexapole} at the cost of fast depletion of the dispenser reservoir. To enhance the lifetime, dispensers are typically operated at reduced flux.~\cite{catani2006intense, holtkemeier20112d}~In such cases, loading from dispensers can be enhanced by placing them in close proximity to the trapping region, such that atoms are directly captured from the atom jet. While such setups have been reported, they typically require complex, customized mounting structures to ensure thermal and electrical isolation of the dispensers.~\cite{catani2006intense, dorscher2013creation,kreyer2017cyl}~For the particular case of cesium (Cs), a 2D MOT based on dispensers has not been reported yet.

\begin{figure} 
    \centering
    \includegraphics[width = 8.6 cm]{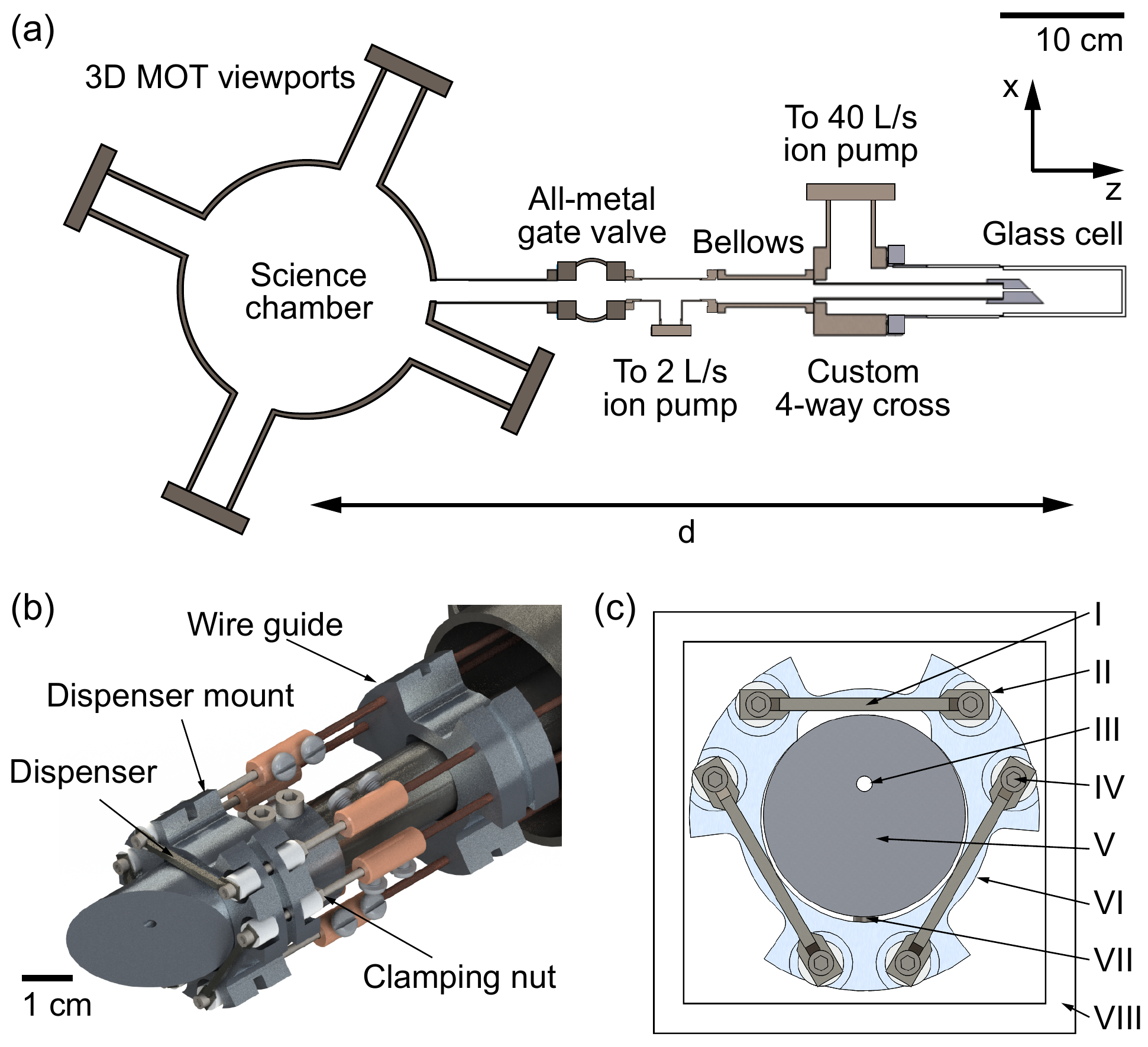}\\
    \caption{Vacuum setup of the 2D MOT. (a) Schematic of the vacuum system. The 2D MOT is formed in a glass cell. It loads a 3D MOT in the science chamber (the wire guide and dispenser mount are omitted for clarity). (b) Rendering of the dispenser mount and 45$\degree$ mirror assembly showing electrical connections and wire guide. (c) Dispenser mount: (I) dispenser; (II) insulating ceramic beads; (III) escape hole; (IV) screws holding dispensers in place and making the electrical connections; (V) 45$\degree$ mirror;  (VI) custom aluminum dispenser mount; (VII) set screw completing the three point contact between the dispenser mount and the mirror; (VIII)  glass cell.}
    \label{fig:2D_MOT_cartoon_Full}
\end{figure}

%In this letter, we present a 2D MOT design that integrates dispensers into a compact glass cell and performs on par with more complex existing designs. \green{We demonstrate a peak flux of $4 \times 10^8$ atoms/s, which is comparable to cesium Zeeman slowers that generate about $8 \times 10^8$ atoms/s,~\cite{hung2011situ} while at the same time providing technical and operational advantages as discussed below.} Aided by a numeric model, we also gain insights into the performance of our setup. Our cesium 2D MOT is part of a multi-species apparatus for the creation of ultracold sodium-cesium molecules.~\cite{lam2020double}

%In this letter, we present a Cs 2D MOT that integrates dispensers into a glass cell in close proximity and direct line of sight of the trapping trapping region. Our setup has a peak flux of $4 \times 10^8$ atoms/s, which is comparable to Cs Zeeman slowers that generate about $8 \times 10^8$ atoms/s, while the vapor pressure in the glass cell is in the $10^{-9}$ torr range. The setup is robust and easy to build. In our lab, it is used in a multi-species apparatus for the creation of ultracold sodium-cesium molecules~\cite{lam2020double} and has operated maintenance-free since construction.

In this letter, we present a compact 2D MOT for Cs that integrates dispensers into a glass cell in close proximity to the trapping region. The setup is robust and easy to build. The vapor pressure in the glass cell is in the $10^{-9}$ torr range. We achieve a peak flux of $4 \times 10^8$ atoms/s, which is comparable to the flux of Cs Zeeman slowers that generate about $8 \times 10^8$ atoms/s.\cite{hung2011situ} In our lab, the 2D MOT is used in a multi-species apparatus for the creation of ultracold sodium-cesium molecules~\cite{lam2020double} and has operated maintenance-free since construction.

\section{Experimental Setup}

\subsection{Vacuum System}

Fig.~\ref{fig:2D_MOT_cartoon_Full}(a) shows a schematic of the experimental apparatus into which the 2D MOT is integrated. It consists of two UHV regions: one for the 2D MOT glass cell,~\cite{cell} and one for the 3D MOT in the science chamber. The 2D MOT and the 3D MOT are separated by $d=61(6)$~cm, where the uncertainty reflects the extent of the elongated 2D MOT trapping region. The vacuum in the $4 \times 4 \times 10$ cm glass cell is maintained by a 40~L/s ion pump connected to the bottom of a custom 4-way cross, which is designed to minimize the distance between the glass cell and the science chamber. The glass cell is connected to the science chamber via a narrow cylindrical hole (1.5~mm diameter, 20~mm length) in a 45$\degree$ in-vacuum mirror. The hole serves as an escape hole for the atomic beam and as a differential pumping tube. The 45$\degree$ aluminum mirror reflects in and out the counter-push and push laser beams, respectively. The pressure differential between the glass cell and science chamber is $10^{-3}$, enabling a pressure of $\ll 10^{-11}$ Torr in the science chamber at $3 \times 10^{-9}$ Torr in the 2D MOT region. The escape hole is vertically offset by 3~mm to take into account the gravitational drop of the atomic beam as it travels towards the science chamber. At the output side of the 2D MOT, a 2~L/s ion pump ensures UHV in the intermediate region after the 2D MOT is attached to the science chamber and before the gate valve to the main chamber is opened.

\subsection{Dispenser Mounting}

Three dispensers,~\cite{dispenser} each containing 3.9 mg of cesium, are mounted in a compact assembly in close proximity to the trapping region. They are oriented around the 45$\degree$ mirror such that the jet of atoms from the dispensers points axially towards the 2D MOT glass cell (see  Fig.~\ref{fig:2D_MOT_cartoon_Full}(b)). Only one dispenser is operated at a time; the remaining two are installed as spares. Most components of the dispenser assembly are commercially available. The only custom parts are a dispenser mount and a wire guide, both of which have loose machining tolerances.~\cite{drawingFootnote} The wires connected to the dispensers are threaded through ceramic beads~\cite{beads} for electrical and thermal isolation. The dispensers are fixed to the mount by 1 inch long \#0-80 screws, which run through holes punched into the dispenser leads and are clamped to the ceramic beads by nuts (see Fig.~\ref{fig:2D_MOT_cartoon_Full}(b)).  Care is taken to ensure that the screws are not in electrical contact with the dispenser mount.  The screws are connected to 16 gauge, kapton-coated wire via inline barrel connectors,~\cite{barrel} which run through the custom 4-way cross to electrical feedthroughs~\cite{power} (not shown). A 3-point contact with one set screw on the bottom (see  Fig.~\ref{fig:2D_MOT_cartoon_Full}(c)) secures the dispenser mount to the 45$\degree$ mirror, which leaves sufficient gaps to prevent virtual vacuum leaks.  Large cutouts on the sides of the dispenser mount and the wire guide enhance vacuum conductance. Since the current-carrying components heat up during operation, care is taken to avoid any contact with the glass cell. %The wire guide reduces strain on the electrical connections.

\begin{figure}
    \centering
    \includegraphics[width = 8.6 cm]{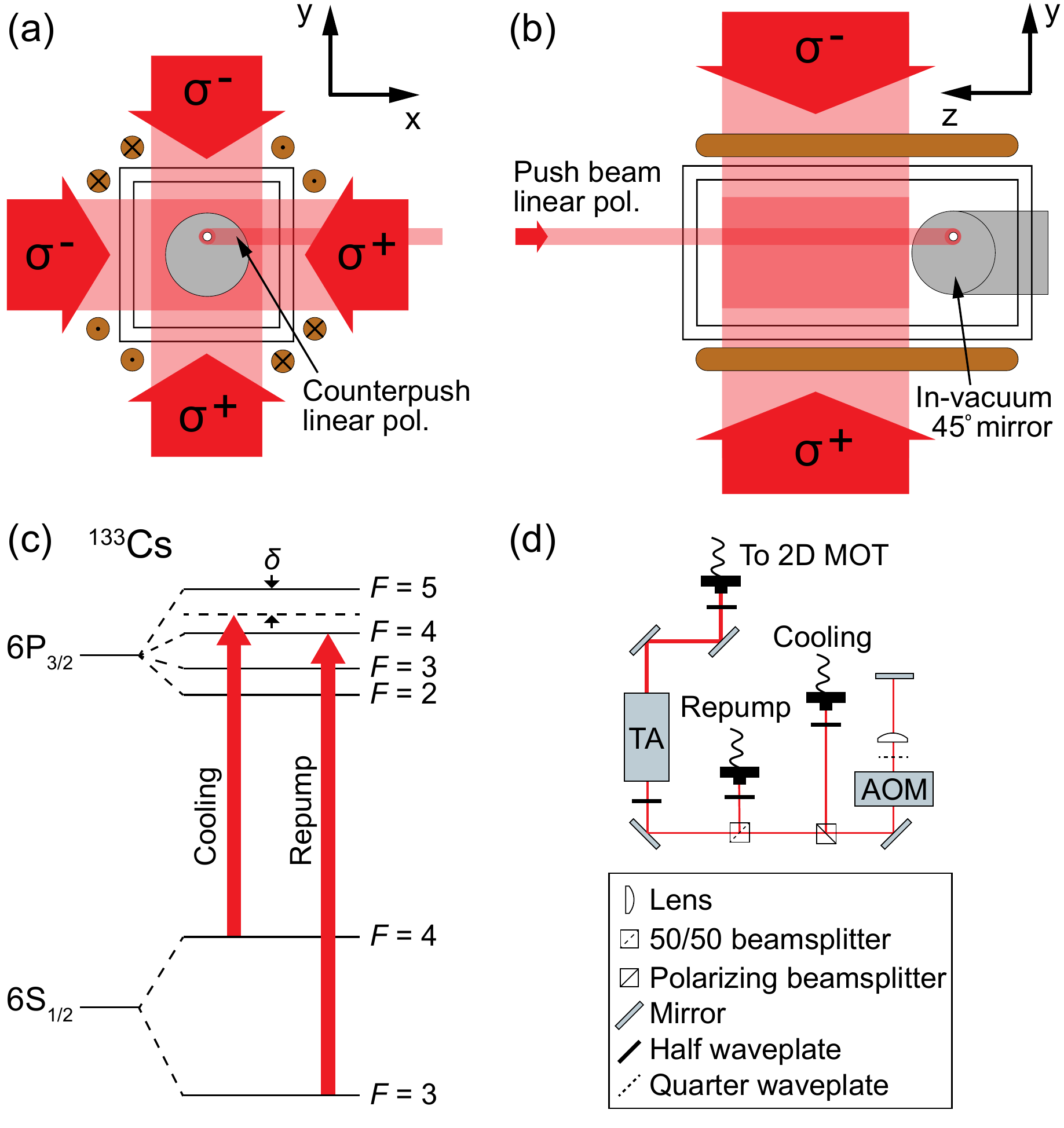}\\
    \caption{Laser setup of the 2D MOT.  (a) Schematic rear view of the glass cell.  Large red arrows represent the MOT beams, where $\sigma^{+}$ and $\sigma^{-}$ indicate the polarizations. In the center of the glass cell is the 45$\degree$ in-vacuum mirror.  The push beam, counter-push beam and the MOT are aligned with the hole in the mirror.  The magnetic fields for the 2D MOT are generated by two pairs of anti-Helmholtz coils, with cross-sections shown as copper-colored circles. Gravity points in the -y direction. (b) Side view of the glass cell. (c) Cesium energy level diagram.  Shown are the cooling laser, with detuning $\delta$ from the $F = 4 \rightarrow F^{\prime} = 5$ transition, and the repump laser, on resonance with the $F = 3 \rightarrow F^{\prime} = 4$ transition. (d) Setup for combining repump and cooling light at a power ratio of 10:1 before a tapered amplifier. The frequency of the cooling light is tuned independently of the repump light using a double-pass AOM.~\cite{detuningFootnote}
    }
    \label{fig:Cs_levels}
\end{figure}

\begin{table}
    \caption{Parameters of 2D MOT laser beams.}
    \label{tab:2D_MOT_beams}
    \begin{ruledtabular}
    \begin{tabular}{cccc}
         Beam & Power (mW) & Polarization & $1/e^2$ Radius (mm) \\
         \hline
         Horizontal & 150 & Right-circular &  8 $\times$ 34 \\
         Vertical & 150 & Left-circular &  8 $\times$ 34 \\
         Push & 3 & Vertical & 2.2 \\
         Counter-push & 0.3 & Horizontal & 2.2
    \end{tabular}
    \end{ruledtabular}
\end{table}

\subsection{2D MOT Laser Setup}

The laser beams and magnetic field coils of the 2D MOT are shown in Figs.~\ref{fig:Cs_levels}(a) and \ref{fig:Cs_levels}(b). We use elongated beams with an aspect ratio of 4:1. The 45$\degree$ in-vacuum mirror allows independent push and counter-push beams to be aligned collinearly to the escape hole, forming a 2D-plus MOT.~\cite{dieckmann1998two,schoser2002intense,chaudhuri2006realization,catani2006intense,ridinger2011large} The mirror is made from electro-polished aluminum and has a reflectance of about 60\%. In our implementation, we fix the power balance between the push and counter-push beams at approximately 10:1. In the region where the push beam reflects off the 45$\degree$ mirror, a transverse force arises, leading to an unwanted deflection of the atomic beam at the location of the escape hole. The power ratio of the push and counter-push beams is empirically optimized to minimize this deflection and allow for maximum flux through the escape hole.

In the remainder of the paper we discuss the implementation of the 2D-plus MOT for the specific case of cesium-133 atoms. The relevant energy levels for laser cooling, using the Cs D$_2$ transition~\cite{steck2003cesium} at $\lambda = 852$~nm, are shown in Fig.~\ref{fig:Cs_levels}(c). The \textit{cooling} laser operates on the cycling transition $(6^2\mathrm{S}_{1/2})$ $F = 4 \rightarrow\,(6^2\mathrm{P}_{3/2})\,F^{\prime} = 5$; the \textit{repump} laser operates on the $F = 3 \rightarrow F^{\prime} = 4$ transition and closes the cooling cycle. The saturation intensity of the Cs D$_2$ line is $I_{\rm sat} = 1.1 $~ mW/cm.$^2$

The light for the 2D-plus MOT is provided by two extended cavity diode lasers  (ECDLs) that are independently locked to cesium vapor cells via saturated absorption spectroscopy. A double-pass acousto-optic modulator (AOM) in the path of the cooling laser allows control of the detuning $\delta$ (see Fig.~\ref{fig:Cs_levels} (c)). The two ECDLs simultaneously seed a tapered amplifier (TA) to produce a single beam containing both cooling and repump light. The TA provides a laser power of up to 300~mW. A sketch of the TA system is shown in Fig.~\ref{fig:Cs_levels}(d). 

The light is distributed into two retro-reflected MOT beams, the push beam, and the counter-push beam, as illustrated in Figs.~\ref{fig:Cs_levels}(a) and~\ref{fig:Cs_levels}(b). The parameters of the beams are detailed in Table~\ref{tab:2D_MOT_beams}. The elongated MOT beams are shaped by a lens system consisting of an asphere and two perpendicularly oriented cylindrical lenses.~\cite{kreyer2017cyl} After passing through the glass cell, each MOT beam is focused onto a mirror that retro-reflects the beam in a cat's eye configuration through a quarter-waveplate to reverse the circular polarization. The incoming and retro-reflected MOT beams must be precisely overlapped to avoid unwanted transverse optical forces. 

The quadrupole magnetic field for the 2D MOT is generated using two racetrack-shaped coil pairs (inner diameters 11.5~cm x 2.3~cm with a minimum coil separation of 5.7~cm), which produce gradients in the $x$ and $y$ directions of up to 24~G/cm in the center of the glass cell at a 5~A current.~\cite{gradientFootnote}  The coils have 48 windings of 16 gauge magnet wire and are mounted inside of a 3D-printed enclosure around the glass cell. The current through each coil can be tuned independently, which allows for complete control over the magnetic field zero in the $x$ and $y$ directions and the strength of the gradient. In practice, we fix the current in one coil on each axis and use the currents in the remaining two coils to align the trapping region with the escape hole.

\section{Numeric 2D MOT Model}

In addition to the experimental characterization discussed below, we model the trapping performance of the 2D MOT using a numeric Monte Carlo simulation.

We assume the 2D MOT is loaded from the background Cs vapor in the glass cell, where the initial velocity distribution of the atoms is assumed to correspond to a Maxwell-Boltzmann distribution. For an atom randomly picked from this statistical ensemble, we sum up the Doppler forces from each laser beam. The acceleration exerted on an atom by a laser beam of wavevector $\mathbf{k}$ is given by
\begin{equation}
\mathbf{a} =  \frac{\hbar \mathbf{k}}{m} \frac{\Gamma}{2} \frac{S_0}{ 1 + S_0 + (2 \Delta / \Gamma)^2},
\label{eq:acceleration}
\end{equation}
where $\Gamma$ is the atomic transition linewidth, $m$ the atomic mass, and $S_0 = I/ I_{\rm sat}$ the resonant saturation parameter with the atomic saturation intensity $I_{\rm sat}$ and the laser intensity $I$, which takes into account the Gaussian profile of the laser beams.  The detuning $\Delta$ of the laser, adjusted for the Doppler and Zeeman effects, is
\begin{equation}
\Delta = \delta - \mathbf{v} \cdot \mathbf{k} - \mu_B B(\mathbf{r})/\hbar.
\label{eq:detuning}
\end{equation}
Here, $\delta$ is the laser detuning from the bare atomic resonance, $\mathbf{v}$ is the velocity of the atom, $B(\mathbf{r})$ is the magnitude of the magnetic field at the location of the atom, $\mathbf{r}$, and $\mathbf{k}$ the wavevector pointing in the direction of laser propagation with $|\mathbf{k}|=2 \pi / \lambda$. Using this force model, we run a classical Monte Carlo trajectory simulation for our 2D MOT beam geometry.

In order to quantify trapping success in the 3D MOT, the simulation first determines if an atom is captured in the 2D MOT, then if it reaches the 3D MOT region in the science chamber, and finally if it is captured by the 3D MOT. The Cs atom velocities are sampled from a Maxwell-Boltzmann distribution at room temperature ($T=300\,$K),~\cite{tempFootnote} and the positions and directions are chosen at random. The simulation of a trajectory terminates either when an atom hits the cell walls, or when it is captured by the 3D MOT. By running 300,000 trajectories,~\cite{simFootnote} we determine the capture efficiency, $\alpha$, which is the probability of an atom of the initial Maxwell Boltzmann distributions to be captured in the 3D MOT. We determine optimal parameters for 2D MOT operation by maximizing the capture efficiency in the 3D MOT. Under optimized conditions, we determine an $\alpha$ of $6 \times 10^{-5}$. 

\begin{figure} [tb]
    \centering
    \includegraphics[width = 8.6 cm]{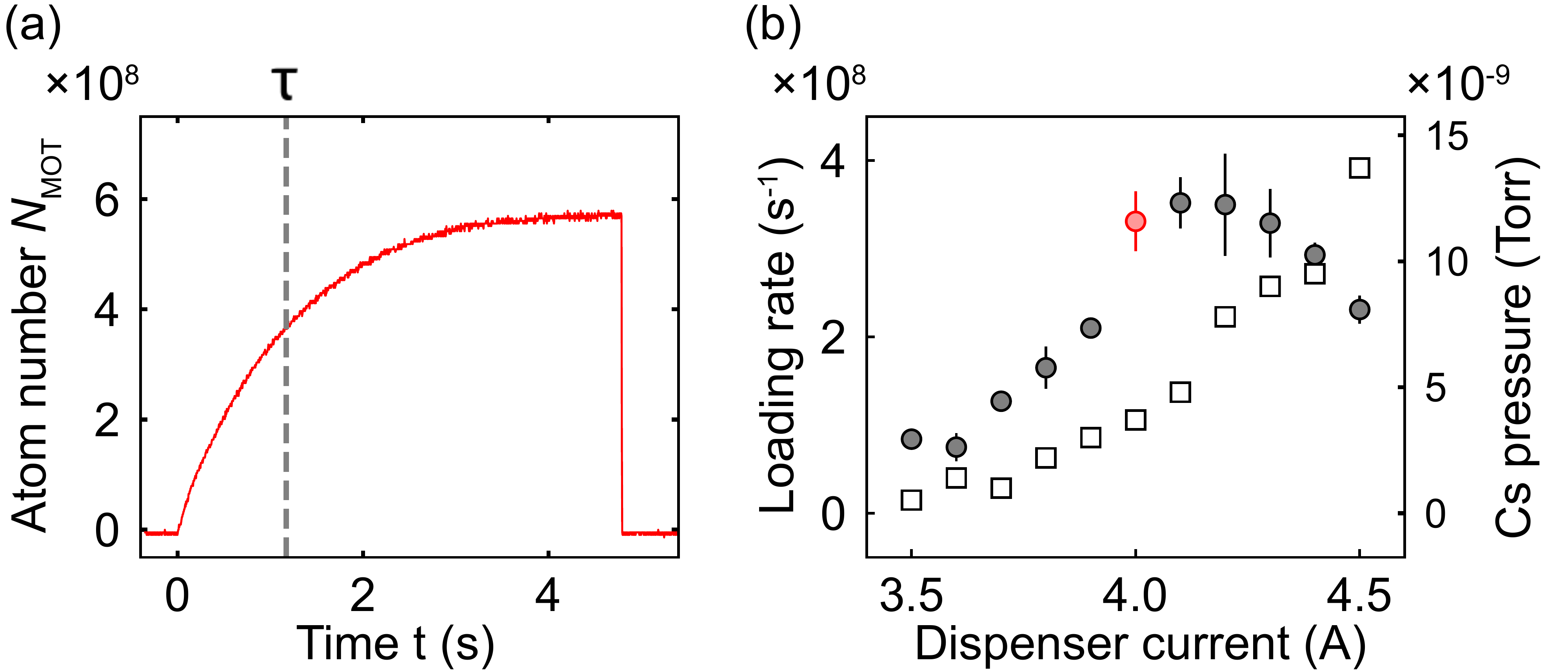}\\
    \caption{Loading of the 3D MOT. (a) Atom number in the 3D MOT as a function of time. The gray dashed line indicates the loading time constant $\tau$. (b) 3D MOT loading rate (gray circles) and Cs vapor pressure (white squares) as a function of dispenser current. The dispenser is run continuously during measurement. Each data point for the loading rate is the mean of two measurements. The red data point indicates the dispenser current used in the remainder of the paper.
    }
    \label{fig:Dispenser_Optimization}
\end{figure}

Using the simulated capture efficiency, we also estimate the loading rate in the 3D MOT. Naturally, this estimate only has order-of-magnitude character as it involves experimental parameters that are hard to determine with high accuracy. We estimate the loading rate~\cite{schoser2002intense} $L$, using $L = \alpha N_\mathrm{cell} / \tau_\mathrm{capture}$, where $\alpha$ is the capture efficiency provided by the simulation, $N_\mathrm{cell} = 9 (4) \times 10^9$, is the number of Cs atoms in the cell determined by a fluorescence measurement (see section~\ref{sec:expdata}), and the 2D MOT loading time constant, $\tau_\mathrm{capture} = 11(5)$~ms, is the time it takes for an atom to be captured in the 2D MOT as determined by the simulation. The error bar reflects the uncertainty in determining, when exactly an atom has entered the 2D MOT in the simulation. We obtain an estimated loading rate of $5(4)\times 10^7$~atoms/s.

In the following section we compare simulation results for varying 2D MOT parameters (laser detuning, intensity, magnetic field gradient, etc) to experimental data.

%\begin{table} [b]
%\caption{Cs 3D MOT loading simulation results.  Capture efficiency %comes from the Monte Carlo simulation, while the loading rates are %calculated from Eq.~(3) and Eq.~(4). % The uncertainties in the loading %rates are from uncertainties in both in the number of atoms in the cell %and the two lifetimes.
%}
%    \label{tab:sim_results}
%    \begin{ruledtabular}
%    \begin{tabular}{ccc}
         %Model & Capture efficiency, $\alpha$ & Loading rate, $L$ %(atoms/s)\\
         %\hline
         %Background vapor & $ 6 \times 10^{-5}$ & $5 (4) \times10^{7}$ %\\
%         Dispenser jet & $3\times 10^{-4}$ & $5 (4) \times 10^{8}$
%    \end{tabular}
%    \end{ruledtabular}
%\end{table}

\section{Experimental Data} 
\label{sec:expdata}

In the experiment, we use the 3D MOT loading rate to quantify and optimize the performance of the 2D MOT. The 3D MOT loading rate is a conservative lower bound for the 2D MOT flux.
We measure it by fitting loading curves such as the one shown in Fig.~\ref{fig:Dispenser_Optimization}(a). We model the time-dependent atom number $N_\mathrm{MOT}$ during MOT loading with a first order differential equation $\dot{N}_\mathrm{MOT} = L - N_\mathrm{MOT}/\tau$, where $\tau$ is the time constant of the 3D MOT loading process. We fit experimental MOT loading curves as shown in Fig.~\ref{fig:Dispenser_Optimization}(a) to the solution, %of Eq.~(\ref{Eqn:MOT_number})
which has the form  $N_\mathrm{MOT}(t) = ( 1 - e^{-t/\tau}) L \tau$, and determine $L$ and $\tau$. At peak performance, we find a loading time constant of $\tau \sim 1$~s and a peak atom number of $5 \times 10^8$ Cs atoms in the 3D MOT, as shown in Fig.~\ref{fig:Dispenser_Optimization}(a).

We experimentally optimize the relevant parameters of the 2D MOT one-by-one, while leaving all other parameters unchanged. First, we optimize the dispenser current, as shown in Fig.~\ref{fig:Dispenser_Optimization}(b). As expected, higher dispenser current leads to a monotonic increase of Cs vapor pressure in the 2D MOT, which we confirm by measuring the density of Cs atoms in the glass cell via atomic fluorescence detection and applying the ideal gas law. However, higher dispenser current does not automatically lead to higher atom flux out of the 2D MOT. The loading rate peaks at $4 \times 10^8$~atoms/s at a dispenser current of 4.2~A. We attribute this behavior to collisional loss as hot atoms of the dispenser jet increasingly collide with slow atoms of the cold atomic beam. The subsequent measurements are performed at a dispenser current of 4~A, chosen as a compromise between reaching high flux and economic use of the dispenser.

\begin{figure} [t]
    \centering
    \includegraphics[width = 8.6 cm]{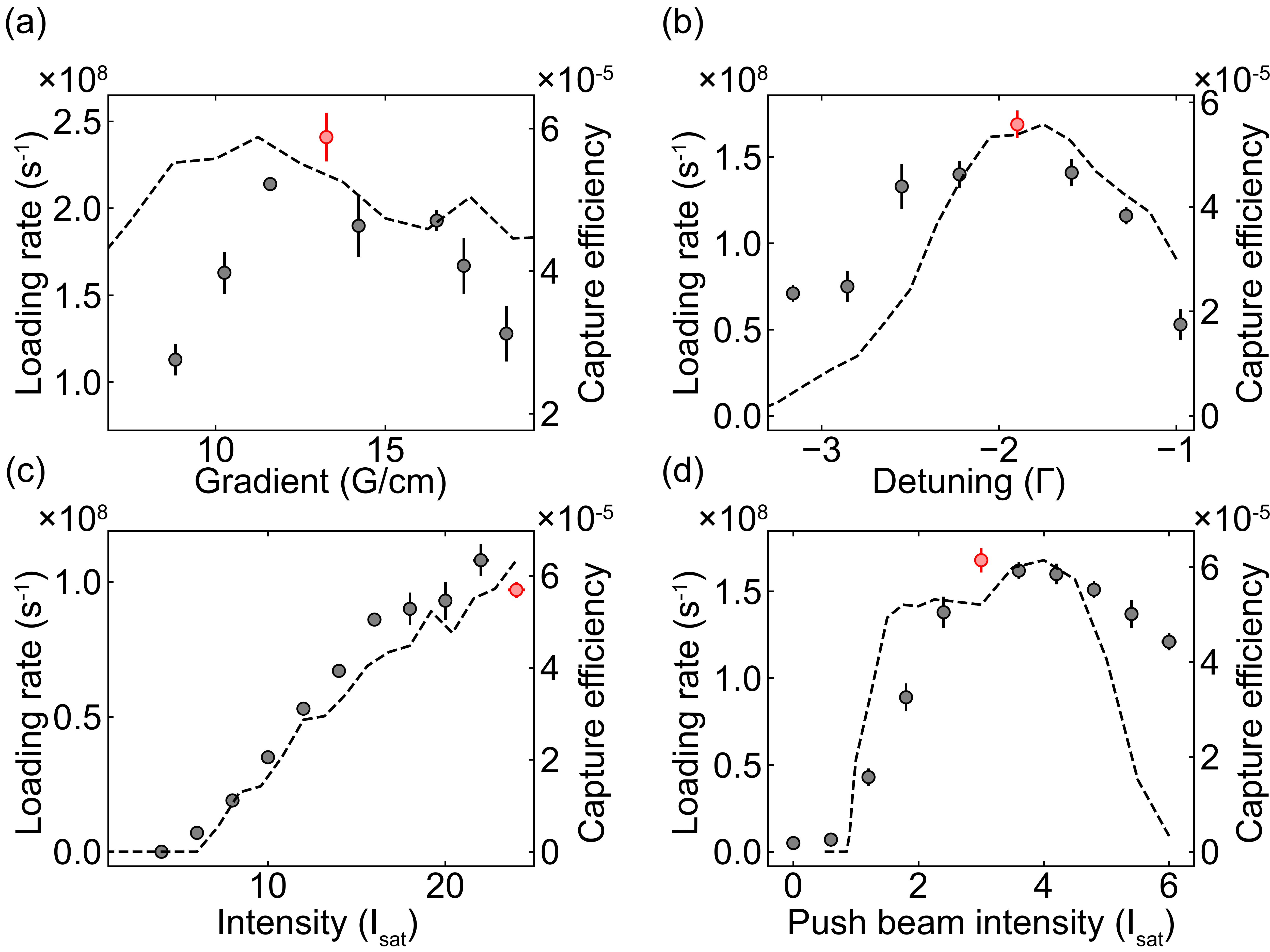}\\
    \caption{Optimizing the 2D MOT performance. The 3D MOT loading rate is monitored, while varying the 2D MOT (a) B-field gradient, (b) cooling laser detuning, (c) cooling beam intensity, and (d) push beam intensity. For each parameter scan, the other parameters are fixed at the values highlighted as red data points. Each data point is the mean of three measurements. Error-bars correspond to the standard deviation of the mean. Dashed lines show the capture efficiency calculated from the numeric Monte Carlo simulation, scaled to match the experimental peak values in each panel.  %We plot the capture efficiency scaled so that the loading rate maximum and the capture efficiency maximum are the same.  Capture efficiency is used because it does not suffer from the large systematic uncertainties in the calculated loading rate while accurately displaying the trend.
    }
    \label{fig:2D_MOT_Optimization}
\end{figure}

Next, we optimize the magnetic field gradient and the detuning of the cooling laser, as shown in Figs.~\ref{fig:2D_MOT_Optimization}(a) and (b). The data is plotted along with the results of the numeric Monte Carlo model. We find that the loading rate peaks at a gradient of 13(2)~G/cm and detuning $\delta = - 1.9\, \Gamma$, where $\Gamma = 2 \pi \times 5.22$~MHz is the natural linewidth of the Cs D$_2$ transition.~\cite{steck2003cesium} Finally, we optimize the loading rate as a function of 2D MOT beam intensity and push beam intensity, as shown in Figs.~\ref{fig:2D_MOT_Optimization}(c) and (d), while keeping the push-to-counterpush beam balance fixed at 10:1. The simulation data in Fig.~\ref{fig:2D_MOT_Optimization} shows the qualitative trends for the various 2D MOT parameters.

\begin{figure}
    \centering
    \includegraphics[width = 8.6 cm]{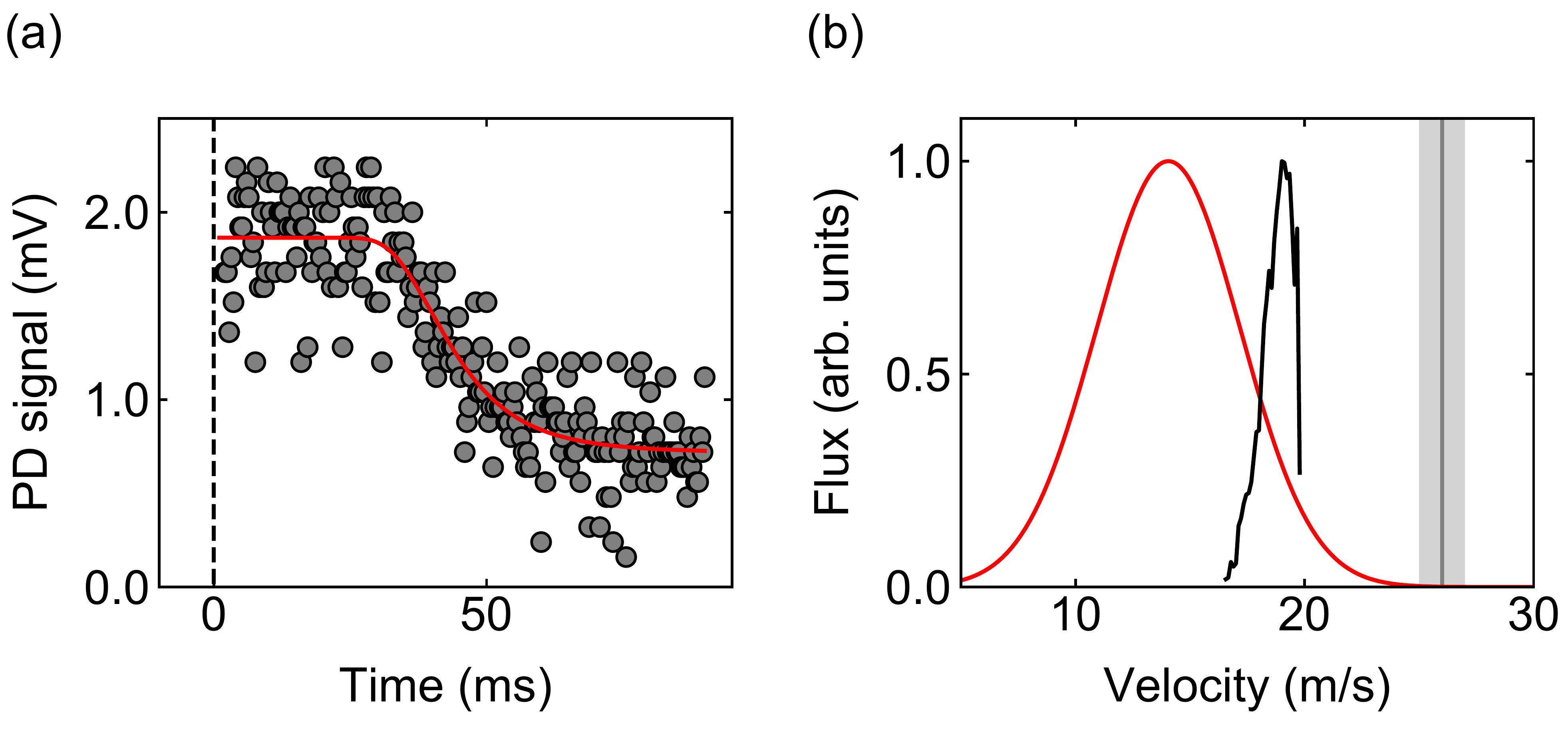}
    \caption{Measuring the 2D MOT velocity distribution at a push beam intensity of 3\,$I_{\mathrm{sat}}$.  (a) Raw fluorescence measurement from time-of-flight measurement (gray points) and fit (red) using Eq.~(\ref{eq:maxboltz}). (b) Corresponding 1D Maxwell-Boltzmann velocity distribution (red).      The vertical gray line is the calculated 3D MOT capture velocity, 26(1)~m/s, with the shaded region representing one standard deviation of uncertainty. The black curve is the calculated velocity distribution from the numeric model. Fitting this curve to a Gaussian, we obtain a center velocity of 19(1)~m/s and a width of 0.8(1)~m/s.}
    \label{fig:2D_MOT_velocity}
\end{figure}

Both the number and the velocity distribution of atoms emerging from the 2D MOT determine the capture success in the 3D MOT. We measure the velocity distribution using a time-of-flight method~\cite{holtkemeier20112d} with the 3D MOT magnetic field coils turned off and blocking one of the six 3D MOT beams to prevent molasses formation.
At $t=0$, the 2D MOT beams are blocked with a mechanical shutter. Then, the temporal decay of the atomic flux from the 2D MOT is detected by measuring the atomic fluorescence in the region of the 3D MOT, as shown in Fig.~\ref{fig:2D_MOT_velocity}(a). To analyze the data, we make the approximation that the atomic beam follows a 1D Maxwell-Boltzmann distribution centered at a non-zero velocity, $v_0$.  We model the decay of atom number in the interaction region as 
\begin{equation}
    \tilde{N}(t) = \tilde{N}_{\rm total} \left ( 1 - \frac{1}{\sqrt{\pi} v_{\rm th}}\int_{d / t}^{\infty} {\rm d} v  \:  e^{ -(v - v_0)^2/v_{\rm th}^2} \right),
    \label{eq:pdint}
\end{equation}
where $v_{\rm th} = \sqrt{2 k_{\rm B} T / m}$ is the width of the thermal distribution, and $d$ is the distance between the 2D MOT and 3D MOT.  The integral term accounts for the atoms that have passed the interaction region at time $t$.  Integrating \eqref{eq:pdint}, we find that the photodiode signal is proportional to
\begin{equation}
    {\rm erf}\left(\frac{d / t - v_0}{ v_{\rm th}}\right).
    \label{eq:maxboltz}
\end{equation}
 We fit Eq.~(\ref{eq:maxboltz}) to the time-of-flight data and determine that the corresponding 1D Maxwell-Boltzmann distribution has a mean velocity of $v_0 = 14(1)$~m/s and a width of $v_{\mathrm{th}} = 5(1)$~m/s (see Fig.~\ref{fig:2D_MOT_velocity}(b)). Given that the capture velocity in the 3D MOT is about $26$~m/s, we expect that the majority of atoms emerging from the 2D MOT is captured.
 
\section{Discussion}\label{sec:discussion}

We compare the measured experimental data to the numeric simulation. Qualitatively, the numeric results capture the trends of the various parameter dependencies, as shown in Fig.~\ref{fig:2D_MOT_Optimization}. The discrepancy in the optimal gradient in  Fig.~\ref{fig:2D_MOT_Optimization}(a) is within the 15\% uncertainty for the determination of the magnetic field gradient in the experimental setup. The measured velocity distribution in Fig.~\ref{fig:2D_MOT_velocity}(b) has a lower mean value and larger width than the numeric simulation. We attribute this difference to diffraction and deformation when the push beam passes through the narrow escape hole, reducing its active area which is not accounted for in the simulation. The slow atomic beam produced by the 2D MOT has a finite transverse velocity. With a smaller active area, atoms spend less time in the push beam and the total longitudinal acceleration from the push beam is reduced. This unintended effect turns out to benefit the performance as the mean velocity of the atomic beam stays significantly below the capture velocity of the 3D MOT. The wider velocity profile also explains the smoother rise of the measured loading rate versus push beam intensity, as shown in Fig.~\ref{fig:2D_MOT_Optimization}(d). Quantitatively, the measured loading rate of up to $4 \times 10^8$ atoms/s is a factor of ten larger than the estimate from the numeric model. Likely, the model underestimates the capture efficiency, as it assumes that atoms hitting the glass cell wall stick to it. Given that there is no metal buildup on the glass cell, we know that atoms can desorb and re-enter the cooling dynamics.

%we find an order of magnitude difference between the simulation and the measured loading rate. This is likely due to our assumption of the Cs atoms taken to be "lost" when it hits a glass surface in the model, which we know cannot be true because there is no Cs metal buildup on the cell.

%the simulation results combined with our measurements, see Table~\ref{tab:sim_results}, indicate that jet-loading leads to ten-times higher loading rates in the 3D MOT over background vapor. In practice, the dispenser jet also creates background vapor in the glass cell. Therefore, we note that both loading mechanisms contribute to the measured loading rate.

Achieving this relatively high atom flux at a relatively low vapor pressure in the 2D MOT cell ensures moderate maintenance requirements. First, the setup does not run into the common issue of ion pump poisoning. Second, the setup can be used for a long time without refilling the atom sources. So far, we have operated the setup with one dispenser at a current of $4.0\,$A for more than 20 months without notable degradation of atom flux. With three dispensers in place, the setup will be maintenance-free for at least 5 years. The lifetime can be further extended by using dispensers with an increased filling. Third, vacuum pressures are almost two orders of magnitude lower than for 2D MOTs based on effusive ovens.~\cite{huang2016intense,dieckmann1998two,fang2015compact} This simplifies differential pumping between the 2D MOT and an attached UHV chamber.%Jet-loading allows for highly efficient use of the dispensers.

%An important metric for a dispenser-based 2D MOT setup is the dispenser lifetime. So far, we have operated the setup with one dispenser at a current of $4.0\,$A for more than one year without notable degradation of the atom flux. From this we expect a lifetime per dispenser of well above 1 year. With three dispensers in place, the setup will be maintenance-free for well above 3 years. The lifetime can be further extended by using dispensers with an increased filling. %Jet-loading allows for highly efficient use of the dispensers.

%\begin{figure}
%    \centering
%    \includegraphics[width = 8.6 cm]{Future MOT v2.png}\\
%    \caption{Schematic of a 2D MOT setup with suggested improvements. %(a) and (b) show axial and side views, respectively.}
%    \label{fig:improvements}
%\end{figure}

The 2D MOT has a small system size and low power consumption, as summarized in Table~\ref{tab:sizecostpower}. With straightforward modifications, these metrics can be further improved: (1) Replacement of the 45$\degree$ in-vacuum mirror with a flat, black anodized aluminum surface, as we see no positive effect from the counter-push beam.~\cite{catani2006intense} Its dominant effect is to balance the transverse optical force generated by the push beam bouncing off the mirror, which will not be necessary with a flat absorbent surface. (2) Decreasing the vacuum conductance of the cell to the pumping region, which increases the Cs lifetime in the cell, allowing for a lower dispenser flux to produce the same background Cs vapor pressure. (3) Positioning of the escape hole closer to the 2D MOT region (axially) to shield the slow atomic beam from the dispenser jet. (4) Replacement of the electromagnets with rare-earth permanent magnets to save power and generate larger field gradients. These modifications can further reduce the vacuum volume of the setup and lead to a power consumption of less than one Watt.

Finally, the design can be readily adapted to cool other atomic species. Based on simulations, we expect that for potassium and rubidium flux rates of several billion atoms per second can be achieved. In addition, it is possible to integrate dispensers for multiple atomic species into the same 2D MOT setup, which offers a path to substantially simplify the source part of multi-species ultracold atom setups.

%design offers a path to integrating several atomic species into the same 2D MOT setup, which could substantially simplify the source part of multi-species ultracold atom setups.

\begin{table}
\caption{In-vacuum volume and power consumption of the 2D MOT setup (without laser system).}
    \label{tab:sizecostpower}
    \begin{ruledtabular}
    \begin{tabular}{ccc}
          Item & Volume (cm$^{-3}$) & Power (W) \\
          \hline
         Glass cell & 260  &  \\
         Custom 4-way cross & 200 & \\
         Vacuum hardware & 230 & \\
         Ion pumps & & 0.1 \\
         Power supplies &  & 26.9  \\
         \hline
         Total & 690 & 27.0\\
    \end{tabular}
    \end{ruledtabular}
\end{table}

\section{Conclusion}

We have demonstrated a Cs 2D MOT that is compact and simple to construct without compromising on performance. The observed atomic flux of $4 \times 10^8$~atoms/s is comparable to cesium Zeeman slowers and sufficient for the vast majority of use cases. While higher atomic flux can be achieved in designs operating at elevated vapor pressures~\cite{kellogg2012compact, fang2015compact}, the presented system enables reliable operation with minimal maintenance. Size and power consumption can be further reduced with straightforward modifications for the use in atomic quantum devices with low size, weight, and power (SWaP) requirements, such as field-deployable atom interferometers or space-deployable atomic clocks.

%We have demonstrated a 2D-plus MOT that features a simple integration of cesium dispensers in a compact form factor. The design prioritizes simplicity without compromising on performance. The observed loading rate of $4 \times 10^8$~atoms/s is within a factor of two of cesium Zeeman slowers,~\cite{hung2011situ} while our setup is simpler, smaller, less expensive, and fast to construct. While it is possible to achieve higher loading rates in designs operating at high vapor pressures,~\cite{kellogg2012compact, fang2015compact} \red{the presented system excels in terms of its compact size and reliability with minimal maintenance.} In addition, we believe that straightforward modifications can further reduce the size and power consumption of the setup for the use in robust atomic quantum devices with low size, weight, and power (SWaP) requirements, such as field-deployable atom interferometers or space-deployable atomic clocks. 

%there is a path to further improve the performance and compactify our design for the use in robust atomic quantum devices with low size, weight, and power (SWaP) requirements, such as field-deployable atom interferometers or space-deployable atomic clocks. 

\section*{Acknowledgements}

We thank M.~Kwon for critical reading of the manuscript. This work was supported by an NSF CAREER Award (Award No.~1848466) and a Lenfest Junior Faculty Development Grant from Columbia University. S.W.~acknowledges additional support from the Alfred P.~Sloan Foundation. I.S.~was supported by the Ernest Kempton Adams Fund. C.W.~acknowledges support from the Natural Sciences and Engineering Research Council of Canada (NSERC) and the Chien-Shiung Wu Family Foundation. 

\section*{Data Availability}

The data that support the findings of this study are available from the corresponding author upon reasonable request.

%\bibliography{Literature}

\begin{thebibliography}{37}
\bibitem [{dra()}]{drawingFootnote}%
  \BibitemOpen
  \href@noop {} {}\bibinfo {note} {Drawings will be made available upon
  request.}\BibitemShut {Stop}%
\bibitem [{det()}]{detuningFootnote}%
  \BibitemOpen
  \href@noop {} {}\bibinfo {note} {The detuning, $\delta$, is precisely
  controlled via AOMs and we determine the field-free $F = 4 \rightarrow
  F^{\prime} = 5$ resonance with an accuracy of better than
  0.1~MHz.}\BibitemShut {Stop}%
\bibitem [{gra()}]{gradientFootnote}%
  \BibitemOpen
  \href@noop {} {}\bibinfo {note} {The gradient is calculated by solving the
  Biot-Savart equation using the measured currents and the knowledge of our
  coil geometry. The relative error of the calculated gradients is estimated to
  be on the 15\% level.}\BibitemShut {Stop}%
\bibitem [{tem()}]{tempFootnote}%
  \BibitemOpen
  \href@noop {} {}\bibinfo {note} {Temperature of the Cs vapor was determined
  via absorption spectroscopy and a fit to the resulting Doppler
  profile.}\BibitemShut {Stop}%
\bibitem [{sim()}]{simFootnote}%
  \BibitemOpen
  \href@noop {} {}\bibinfo {note} {Trajectory simulation was done assuming a
  Maxwell-Boltzmann distribution. Only velocities below a cut-off velocity of
  70~m/s were simulated, as capture of an atom above this limit is extremely
  unlikely. Approximately 4000 successful capture events were
  observed.}\BibitemShut {Stop}%
\end{thebibliography}
%merlin.mbs aipnum4-1.bst 2010-07-25 4.21a (PWD, AO, DPC) hacked
%Control: key (0)
%Control: author (8) initials jnrlst
%Control: editor formatted (1) identically to author
%Control: production of article title (0) allowed
%Control: page (1) range
%Control: year (1) truncated
%Control: production of eprint (0) enabled
%

\end{document}